\documentclass[12pt]{iopart}
\usepackage{epsfig,epsf,amsfonts,amscd}
\newcommand{\R}{{\mathbb{R}}}

\begin{document}
\title{Controlling a resonant transmission across the
$\delta'$-potential: the inverse problem }
\author{
A.V. Zolotaryuk and Y. Zolotaryuk}
\address{
Bogolyubov Institute for Theoretical Physics,
National Academy of Sciences of Ukraine,
03680 Kyiv, Ukraine }
\date{\today}

\begin{abstract}
Recently, the non-zero transmission of a quantum particle through the one-dimensional singular potential given in the form of the derivative of Dirac's delta function, 
$\lambda \delta'(x) $, with $\lambda \in \R$, being a potential strength 
 constant, has been discussed by several authors. The transmission occurs
 at certain discrete values of $\lambda$ forming a resonance set 
$\{ \lambda_n\}_{n=1}^\infty$. 
For  $\lambda \notin \{ \lambda_n\}_{n=1}^\infty$ this potential 
has been shown to be a perfectly reflecting wall. However, this 
 resonant transmission takes place only in the case when the 
 regularization of the distribution $\delta'(x) $ is constructed in 
 a specific way. Otherwise, the $\delta'$-potential is fully 
 non-transparent. Moreover, when the transmission is non-zero, 
 the structure of a resonant set depends on a regularizing 
 sequence $\Delta'_\varepsilon(x)$ that tends to $\delta'(x) $ 
 in the sense of distributions as $\varepsilon \to 0$. Therefore, 
 from a practical point of view, it would be interesting to have 
 an inverse solution, i.e. for a given $\bar{\lambda} \in \R$ to 
 construct such a regularizing sequence $\Delta'_\varepsilon(x)$ 
 that the $\delta'$-potential at this value is transparent. If 
 such a procedure is possible, then this value $\bar{\lambda}$ 
 has to belong to a corresponding resonance set.  The present 
 paper is devoted to solving this problem and, as a result, 
 the family of regularizing sequences is constructed by tuning
  adjustable parameters in the equations that provide a resonance
   transmission across the 
$ \delta'$-potential. This construction can be realized if each 
regularizing sequence $\Delta'_\varepsilon(x)$ depends 
on $\lambda \in \R$ and this is a key point of our approach. 
Next, we can solve the inverse problem if the regularization is constructed from rectangles. Since in some cases the renormalization procedure $\Delta'_\varepsilon(x) \to \delta'(x)$ leads to the existence of an effective $\delta$-interaction, it is reasonable from the beginning to consider the linear combination $V(x) = \eta \delta(x) + \lambda \delta'(x)$ with $(\eta, \lambda) \in \R^2$.  
\end{abstract}

\maketitle

\section{Introduction}

The Schr\"{o}dinger operators with singular
zero-range potentials attract a considerable interest beginning from the pioneering work of Berezin and Faddeev \cite{bf}. These operators (for details and references see book~\cite{a-h}) describe point or contact interactions which are widely used in various applications to quantum physics~\cite{do,ll,se1,ppz,sg,kun}. Intuitively, these interactions are understood as sharply localized potentials, exhibiting a number of interesting and intriguing features. Applications of these models to condensed matter physics (see, e.g.,~\cite{ael,ex,es,ces})
 are of particular interest nowadays, mainly  because of the rapid progress in fabricating nanoscale quantum devices, particularly, thin quantum waveguides~\cite{acf,ce}.

In this paper, we consider the one-dimensional Schr\"{o}dinger equation 
\begin{equation}
-\psi''(x) + V(x)\psi(x) =E \psi(x),
\label{1}
\end{equation}
where the prime stands for the differentiation with respect
to the spatial coordinate $x$ and $\psi(x)$ is the wavefunction
for a  particle of mass $m$ and energy $ E $ (we use units in which $\hbar^2/2m =1$). 
Using the regularization 
\begin{equation}
\varepsilon^{-2} {\cal V}(x / \varepsilon) \to \delta'(x)~~
\mbox{as}~~ \varepsilon \to 0
\label{2}
\end{equation}
 (in the sense of distributions), 
\v{S}eba\cite{se2} has rigorously studied equation~(\ref{1}) with the potential $V(x) =\lambda \delta'(x)$, where $\lambda \in \R$ is an interaction strength constant and $\delta'(x)$ is the derivative of Dirac's delta function $\delta(x)$. Here,  ${\cal V}(\xi)$, $\xi \in \R$, is assumed to be a smooth compactly supported function satisfying the conditions 
\begin{equation}
\int_\R {\cal V}(\xi) d\xi =0~~~\mbox{and}~~~
 \int_\R \xi {\cal V}(\xi) d\xi = -1 .
\label{3}
\end{equation}
 As a result, he has obtained in the $\varepsilon \to 0$ limit the direct sum of the free 
Schr\"{o}dinger operators 
$-d^2/dx^2$ on the negative and positive half-axes of $\R$ with Dirichlet boundary conditions at the origin $x=0$.  In physical terms this means that the $\delta'$-barrier is completely nontransparent for quantum particles. The absence of a nontrivial point interaction in the zero-range limit has prompted \v{S}eba~\cite{se2} to introduce and analyze a less singular interaction (with a renormalized interaction strength) through the limit 
\begin{equation}
V(\alpha;x) = \lambda \lim_{\varepsilon \to 0}
{\delta (x +\varepsilon ) - \delta (x -\varepsilon ) \over
2\varepsilon^\alpha }
\label{4}
\end{equation}
with $\alpha \in (0, 1]$. In the limiting case $\alpha =1$, limit (\ref{4}) gives the unrenormalized point dipole interaction $\lambda \delta'(x)$. For this interaction he has proved 
that for $\alpha < 1/2$ limit (\ref{4}) is trivial, i.e. the system behaves as if the point potential is absent, while for $\alpha > 1/2$ the system splits into two independent subsystems separated on the half-axes $(-\infty, 0)$ and $(0, \infty)$. The only non-trivial case has been proved to occur for $\alpha =1/2$, when the point interaction appears to be  $V(1/2; x)=g\delta(x)$ with the effective coupling constant $g = -\lambda^2/2$. Recently, \v{S}eba's approach has been generalized  from one to two \cite{zci} and 
three \cite{z_jpa} dimensions. More precisely, instead of the one power $\alpha$ in (\ref{4}), the regularization procedure using three powers $\mu, \, \nu, \, \tau$ has been developed. As a result, a three-dimensional manifold has been constructed 
in the $\{\mu, \nu, \tau\}$-space that corresponds to the $\delta$-interaction with different values of the effective coupling constant $g$.  

Afterward there were other attempts to regularize the $\delta'$-potential using both local and nonlocal regular functions. However, the calculations of scattering amplitudes in 
\cite{c-g} on a piecewise $\delta'$-like approximating potentials have discovered that at certain discrete values of $\lambda$ the transmission across the $\delta'$-barrier is non-zero. 
   These calculations contradict \v{S}eba's result for $\alpha =1$ and later on Golovaty and Man'ko\cite{gm}  have rigorously proved the existence of non-zero transmission for a wide class of $\delta'$-like regularizing sequences with compact supports. This discrepancy has prompted Golovaty and Hryniv\cite{gh} to revise \v{S}eba's result. They have proved that for $\lambda \delta'$-like potentials with compact supports there exists a sufficiently large set of the coupling constant $\lambda$ at which the norm convergent limit differs from the operator obtained by \v{S}eba.   

In this paper, we develop a regularizing procedure which allows us to get a transparent regime for a given value of $\lambda$. To this end, for any $\mu > 1$, instead of (\ref{2}), we approximate the singular potential $V(x) =\lambda \delta'(x)$ by  
\begin{equation}
V_\varepsilon(\lambda; x) = \lambda \varepsilon^{2(1-\mu)} 
{\cal V}_\varepsilon( \lambda; \varepsilon^{(1-\mu)} x) \to \lambda \delta'(x)
\label{5}
\end{equation}
where the sequence ${\cal V}_\varepsilon( \lambda; \xi)$ depends in general on $\lambda$ and instead of both conditions (\ref{3}) we require the limiting equality
\begin{equation}
\lim_{\varepsilon \to 0} \int_\R \xi 
{\cal V}_\varepsilon (\lambda; \xi) d\xi =-1 .
\label{6}
\end{equation}

As shown previously \cite{zci,z_jpa}, in some cases the barrier-well regularizing sequence leads in the zero-range limit to an additional effective $\delta$-interaction. Therefore, similarly 
to~\cite{gnn}, it is reasonable to add to the $\delta'$-potential a pure $\delta$-potential, so that the total potential $V(x)$ in this paper is assumed to be the sum of Dirac's delta function and its derivative, i.e.
\begin{equation}
V(x) = \eta \delta(x) + \lambda \delta'(x),
\label{7}
\end{equation} 
where $\eta \in \R$ and $ \lambda \in \R\setminus \{0\}$ are strength interaction constants for the $\delta$- and $\delta'$-potentials, respectively. 

Since we do not impose the first constraint of (\ref{3}),  
 we need to define a space of test functions discontinuous 
at the origin. In this way,
we slightly extend the family of point interactions from the standard $\delta'$-potential to a wider class of $\delta'$-like barriers. 
 
\section{Definition of $\delta$- and $\delta'$-like distributions on a space of discontinuous test functions}

The use of distributions on discontinuous test functions has been considered first by Griffiths\cite{gr} and the general theory in this direction has been developed by Kurasov\cite{ku}. For other applications of the distributions for discontinuous test functions see, e.g.,~\cite{gnn,kse,ni,z_pla}. Here, we restrict ourselves only to the class of discontinuous functions the all derivatives of which are continuous. The only purpose of this slight generalization is to avoid the first equality (\ref{3}) and therefore to keep only the constraint (\ref{6}).   

Let $\varphi(x)$ be a test function from the ${\cal D}$ space. Shifting a positive part of this function by a non-zero constant, one can form a space of discontinuous at $x=0$ test functions. To this end, for a fixed $\varsigma \in \R$ we assume 
\begin{equation}
\varphi_\varsigma (x) \doteq \varphi(x) + (\varsigma -1) \varphi(0) \Theta(x) ,
\label{8}
\end{equation} 
where $\Theta(x)$ is the Heaviside function. This function is discontinuous at $x=0$ if $\varsigma \neq 1$, i.e. $\varphi_\varsigma (-0) = \varphi( 0)$ and $ \varphi_\varsigma (+0)  =\varsigma  \varphi( 0)$, while its two-sided derivatives are continuous at 
$x=\pm 0$. 
For each $\varsigma \neq 1 $ we denote the space of these test functions by 
${\cal D}_{\varsigma}$. Clearly, there exists a one-to-one correspondence between the spaces
${\cal D}$ and ${\cal D}_{\varsigma}$ at a given $\varsigma $ and in the particular case $\varsigma = 1$ the space ${\cal D}_{\varsigma}$ coincides with ${\cal D}$. Due to the boundary conditions at $x=0$, one can slightly modify, e.g., the distributions $\delta(x)$ and $\delta'(x)$ as follows: $\delta(x): \varphi_\varsigma(x) \to \varsigma \varphi(0)$ and $\delta'(x): \varphi_\varsigma(x) \to - \varphi'(0)$.

Now our purpose is to construct the regularizing sequences 
$\Delta_\varepsilon(x)$ and $\Delta'_\varepsilon(x)$ (the prime here does not denote the differentiation) such that 
$\Delta_\varepsilon(x) \to \delta(x)$ and 
$\Delta'_\varepsilon(x) \to \delta'(x)$ in the sense of distributions defined above on the space 
${\cal D}_\varsigma $. Introducing the new spatial variable 
$\xi = \varepsilon^{1-\mu}x $ with $\mu >1$,
we define new regularizing functions  ${\cal V}_\varepsilon (\xi)$ and ${\cal V}'_\varepsilon (\xi)$  through the relations 
\begin{equation}
\Delta_\varepsilon(x) = \varepsilon^{1-\mu} 
{\cal V}_\varepsilon (\xi)
~~~\mbox{and}~~~\Delta'_\varepsilon(x) = \varepsilon^{2(1-\mu)} {\cal V}'_\varepsilon (\xi) .
\label{9}
\end{equation}
Again, the prime in ${\cal V}'_\varepsilon (\xi)$, the same 
as in $\Delta'_\varepsilon(x)$, does not mean differentiation.
Integrating these functions over $\R^-$ and $\R^+$ separately and expanding $\varphi_\varsigma (\xi)$ from the left and the right of the origin $x =\pm 0$, we obtain the expansions 
\begin{eqnarray}
&&\langle \Delta_\varepsilon(x) \, |\, \varphi_\varsigma(x) \rangle \doteq 
 \int_\R \Delta_\varepsilon(x) \, \varphi_\varsigma(x) dx 
= \int_\R {\cal V}_\varepsilon ( \xi ) \, 
\varphi_\varsigma \left(\varepsilon^{\mu -1} \xi \right)d\xi 
\nonumber \\
&=& \int_{\R^-} {\cal V}_\varepsilon ( \xi ) 
\left[ \varphi(0) + \varepsilon^{\mu -1} \xi \, \varphi'(0) 
+ \ldots \right]d\xi 
\nonumber \\
&+& \int_{\R^+} {\cal V}_\varepsilon ( \xi ) 
\left[  \varsigma  \, \varphi(0) + 
\varepsilon^{\mu -1} \xi \, \varphi'(0) + \ldots \right]d\xi
\nonumber \\ 
&=& m_{0, \varsigma} (\varepsilon) \, \varsigma \varphi(0) +  \ldots + \varepsilon^{j(\mu-1)} \,  m_{j}(\varepsilon) \, 
\varphi^{(j)}(0) + \ldots ,
\label{10}
\end{eqnarray}
where
\begin{equation}
m_{0, \varsigma} (\varepsilon) \doteq  \varsigma^{-1} \int_{\R^-} 
{\cal V}_\varepsilon ( \xi )d\xi \, \, 
+ \int_{\R^+} {\cal V}_\varepsilon ( \xi )d\xi, ~~
m_j(\varepsilon) \doteq {1 \over j! }
\int_\R \xi^j {\cal V}_\varepsilon ( \xi )d\xi , 
\label{11}
\end{equation}
with $j = 1,2, \ldots $.
As follows from this expansion, for the proper definition of the $\delta (x)$ function, we need to have 
\begin{equation}
\lim_{\varepsilon \to 0}
m_{0, \varsigma}(\varepsilon) =1~~\mbox{and}~~ 
 \lim_{\varepsilon \to 0} m_j(\varepsilon) = \mbox{const} 
\label{12}
\end{equation}
 for all $j = 1,2  \ldots$.

 Similarly, we write
\begin{eqnarray}
 && \langle \Delta'_\varepsilon(x) \, |\, \varphi_\varsigma(x) \rangle \doteq 
\int_\R \Delta'_\varepsilon(x) \, \varphi_\varsigma(x) dx 
= \varepsilon^{1-\mu} \int_\R {\cal V}'_\varepsilon ( \xi ) \, 
\varphi_\varsigma\left(\varepsilon^{\mu -1} \xi \right)d\xi 
 \nonumber \\
&= &  \varepsilon^{1-\mu} \int_{\R^-} {\cal V}'_\varepsilon ( \xi ) 
\left[ \varphi(0) + \varepsilon^{\mu -1} \xi \, \varphi'(0) 
+ \ldots \right]d\xi \, \,
  \nonumber \\
&+ & \varepsilon^{1-\mu} \int_{\R^+} 
{\cal V}'_\varepsilon ( \xi ) 
\left[  \varsigma  \, \varphi(0) + 
\varepsilon^{\mu -1} \xi \, \varphi'(0) + \ldots \right]d\xi
\nonumber \\
 &=&\varepsilon^{1-\mu } \, m'_{0, \varsigma} (\varepsilon)\, \varsigma  \varphi(0)  
+ \ldots + \varepsilon^{(j-1)(\mu-1)} 
\,  m'_j(\varepsilon) \, \varphi^{(j)}(0) + \ldots ,
\label{13}
\end{eqnarray}
where 
\begin{equation}
m'_{0, \varsigma} (\varepsilon) \doteq \varsigma^{-1} \int_{\R^-} 
{\cal V}'_\varepsilon ( \xi )d\xi \, \, 
+  \int_{\R^+} {\cal V}'_\varepsilon ( \xi )d\xi, ~~
m'_j(\varepsilon) \doteq {1 \over j! }
\int_\R \xi^j {\cal V}'_\varepsilon ( \xi )d\xi , 
\label{14}
\end{equation}
with $j =1,2,  \ldots $.
Here, we have to examine the following two possibilities:  
$\Delta'_\varepsilon(x) \to \delta(x)$ and 
$\Delta'_\varepsilon(x) \to \delta'(x)$ on the space of discontinuous test functions from the 
 ${\cal D}_{ \varsigma}$ space. In the former case, we have to satisfy the equations
\begin{equation}
\lim_{\varepsilon \to 0}\varepsilon^{1-\mu}
 m_{0, \varsigma}'(\varepsilon) =1~~ \mbox{and}~~
\lim_{\varepsilon \to 0} m'_j(\varepsilon) =0 
\label{15}
\end{equation} 
for all $j =1, 2, \ldots ,$
 whereas in the latter case, we need to have 
\begin{eqnarray}
m'_{0, \varsigma}(\varepsilon) =0~~\mbox{for any}~ \varepsilon > 0,~~\lim_{\varepsilon \to 0} 
m'_1(\varepsilon) = -1~~ \mbox{and}~~ \lim_{\varepsilon \to 0} m'_j(\varepsilon) = \mbox{const} 
\nonumber \\
\label{16}
\end{eqnarray} 
 for all $j = 2,3  \ldots$.

\section{A rectangular model and its power parametrization }

In this paper, we construct a regularizing sequence for the singular potential 
(\ref{7}) consisting of three adjacent rectangular  barriers/wells.  More precisely, the regularizing sequences $\Delta_{\varepsilon}(x)$ and $\Delta'_{\varepsilon}(x)$ for the rectangular model are specified through piecewise functions as follows:
\begin{equation}
\Delta_{\varepsilon}(x) = \left\{ \begin{array}{ll}
  h  &   \mbox{for}~~ 0  < x < \rho , \\
0 & \mbox{otherwise} 
\end{array} \right. ~\mbox{and}~~
\Delta'_{ \varepsilon}( x) = 
\left\{ \begin{array}{llll}
 h_1   &  \mbox{for}~~ -l < x < 0 ,  \\
h_2 & \mbox{for}~~~ \rho  < x < \rho +r  , \\
  h_3   &  \mbox{for}~~ ~\, 0 < x < \rho ,  \\
 ~  0 & ~~\mbox{otherwise} 
\end{array} \right. 
\label{17}
\end{equation} 
with the constraint $h_1 h_2 <0$ (double-well structure).
We parametrize the rectangular parameters
 $h,h_1, h_2, h_3, l,\rho,r$ by powers as 
\begin{eqnarray}
h_1&=&a_1\varepsilon^{2(1-\mu)} + {c_0 \over c_1}
\varepsilon^{-\mu},~~ h_2= a_2\varepsilon^{2(1-\mu)} -
{c_0 \over c_2} \varepsilon^{-\nu},~~h_3=a_3 
\varepsilon^{2(1-\mu)}, \nonumber \\
 h&=& c_3^{-1}\varepsilon^{-\tau}, ~~
l = c_1 \varepsilon,~~r=
{ c_2 \over \varsigma}\varepsilon^{1-\mu+\nu}, 
~  ~ \rho = c_3 \varepsilon^\tau , 
\label{18}
\end{eqnarray}
where $a_i ~(i=1,2,3)$, $c_j~ (j=0,1,2,3)$, $\varsigma$ are positive numbers, and $\varepsilon$ is a squeezing parameter.

Inserting parametrization (\ref{18}) into (\ref{17}), according to definition (\ref{9}), one can write
\begin{equation}
{\cal V}_\varepsilon (\xi)= \left\{ \begin{array}{ll}
c_3^{-1} \varepsilon^{\mu-1-\tau} &  \mbox{for}~~
  0 < \xi <  c_3 \varepsilon^{1- \mu +\tau} , \\
~~~~0 & \mbox{otherwise} 
\end{array} \right.
\label{19}
\end{equation}
and
 \begin{equation}
{\cal V}'_\varepsilon(\xi) =
\left\{ \begin{array}{lllll}
a_1 + (c_0/c_1)\varepsilon^{\mu-2} & \mbox{for} ~~
 - c_1 \varepsilon^{2-\mu} < \xi < 0 , \\
a_2 -  (c_0/c_2)\varepsilon^{2\mu-2-\nu} &  \mbox{for}~~
   c_3 \varepsilon^{1- \mu +\tau} < \xi < (c_2/\varsigma)
 \varepsilon^{2-2\mu +\nu} \\ & ~~~~~~ + c_3 \varepsilon^{1- \mu +\tau}, \\
~~~~~~a_3 & \mbox{for} ~~ 0 < \xi < c_3 
\varepsilon^{1-\mu +\tau} , \\
~~~~~~~0 & \mbox{otherwise} .
\end{array} \right.
\label{20}
\end{equation}

Obviously, for any $\varepsilon > 0$ we have 
$m_{0, \varsigma} (\varepsilon)= 1$ and the second limit 
(\ref{12}) is easily proved by induction for any positive 
$c_3$ if $\tau \ge \mu -1$. Therefore, in this case 
$\Delta_{\varepsilon}(x) \to \delta(x)$ in the sense of the ${\cal D}'$-distributions.  As mentioned above, one can consider separately the $\delta$- and 
$\delta'$-limits of $\Delta'_\varepsilon (x)$ and this depends on the choice of the constants $a_1, a_2, a_3$. Indeed, if these constants are non-zero, the first limit 
(\ref{15}) leads to 
\begin{equation}
\lim_{\varepsilon \to 0} \left[ \varepsilon^{2(1-\mu)}
\left( {a_1 c_1 \over \varsigma} \varepsilon +
{a_2 c_2 \over \varsigma} \varepsilon^{1-\mu +\nu} +
a_3 c_3 \varepsilon^\tau \right)\right] =1 .
\label{21}
\end{equation}
\begin{figure}
\centerline{\epsfig{file=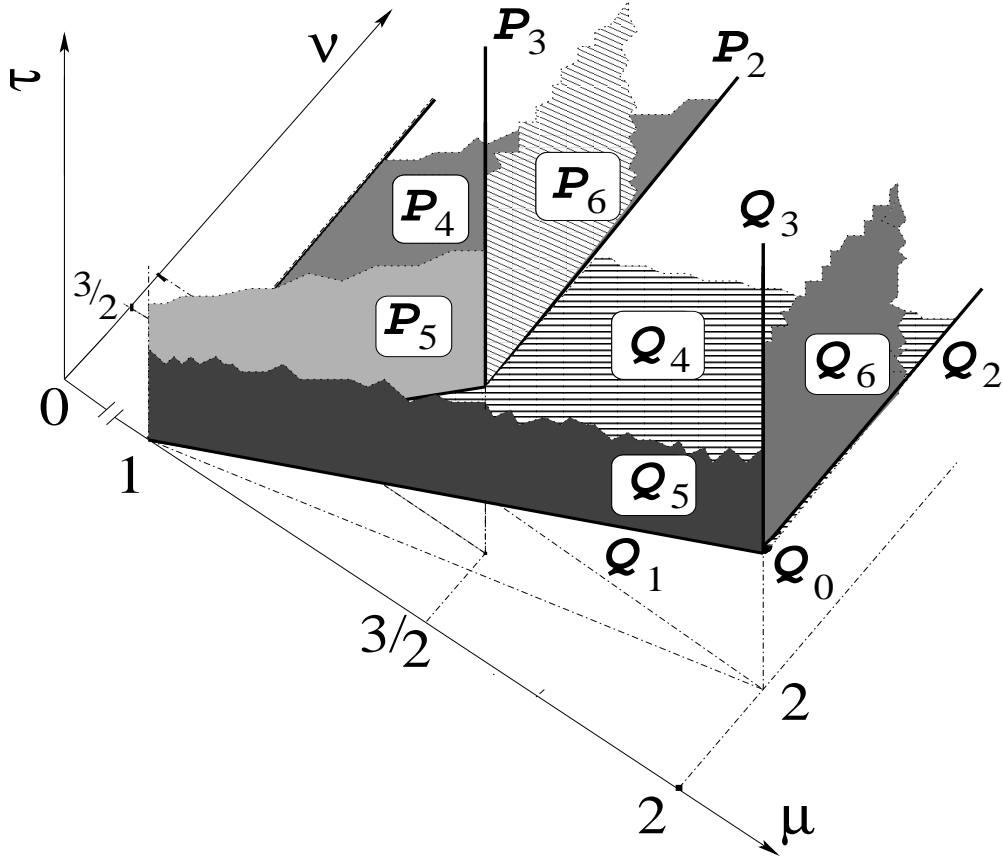,width=5.2 in,
height=4.5 in,angle=0}}
\vspace{2pt}
\caption{
Trihedral surfaces $S_{\delta}$ and $S_{\delta'}$ defined through limits 
(\ref{21}) and (\ref{22}). The surface $S_{\delta}$ is formed by apex $P_0$,
edges $P_1$, $P_2$, $P_3$ and planes $P_{4}$, $P_{5}$, 
$P_{6}$. Similarly, 
the surface $S_{\delta'}$ is formed by apex $Q_0$,
edges $Q_1$, $Q_2$, $Q_3$ and planes $Q_{4}$, $Q_{5}$, 
$Q_{6}$. Notations for apex $P_0$ and edge $P_1$ are omitted. }
\label{fig1}
\end{figure}
This limit determines the trihedral surface $S_\delta$ (see Fig.~1) being a subset of the 
$\{ \mu, \nu, \tau \}$-space with $\nu, \tau \ge \mu -1$. On this surface, the second zero limit (\ref{15}) holds and, using direct calculations, it is proved by induction. The $S_\delta$ surface is formed by the apex $P_0$, the edges $P_1, P_2, P_3$, and the planes 
$P_{4}, P_{5}, P_{6}$, defined in table~\ref{tab:table1}. Limiting 
equation (\ref{21}) results in the constraints on the constants $ a_1, 
\, a_2, \, a_3; \, c_1, \, c_2, \, c_3; \, \varsigma$, depending on the sets $P_j,~j=0,\, 1, \, \ldots \, , 6$. These constraints are summarized in 
table~\ref{tab:table1}. 
\begin{table}
\caption{\label{tab:table1}
 Definition of apex $P_0$, 
edges $P_1, \, P_2, \, P_3$ and planes $P_{4}, \, P_{5}, \, P_{6}$ forming the trihedral surface 
$S_\delta$ together with the constraints on constants  $a_1, 
\, a_2, \, a_3; \, c_1, \, c_2, \, c_3; \, \varsigma$, satisfying the limit (\ref{21}). }
\begin{tabular}{ll}
\hline
\hline
$P_j$-sets    & 
Constraints on $a_1, \, a_2, \, a_3; \, c_1, \, c_2, \, c_3; \, \varsigma$ \\
\hline 
 $P_0 \doteq  \{ \mu = \nu = 3/2, ~ 
 \tau =1 \}$  & $ a_1c_1 + a_2c_2  + a_3c_3
\varsigma =\varsigma  $\\
$P_1  \doteq  \{ 1 < \mu < 3/2, ~ \nu =3(\mu -1),~
\tau =2(\mu -1) \} $ & $ a_2c_2  + a_3c_3
\varsigma =\varsigma    $ \\
 $P_2 \doteq  \{  \mu =3/2, ~ \nu > 3/2,~
\tau = 1\} $ & $ a_1 c_1 + a_3c_3
\varsigma =\varsigma   $ \\
$P_3  \doteq  \{ \mu= \nu = 3/2,~
\tau > 1 \}$ & $  a_1c_1 + a_2c_2 =\varsigma  $   \\
 $P_{4}  \doteq  \{ 1 <  \mu < 3/2, ~ 
\nu > 3(\mu -1) ,~ \tau =2(\mu -1) \} $ & $ a_3c_3=1$\\
$P_{5}  \doteq  \{ 1 < \mu < 3/2,
~ \nu = 3(\mu -1),~
\tau > 2(\mu -1) \} $ & $ a_2c_2= \varsigma  $ \\
 $P_{6} \doteq  \{  \mu = 3/2, ~ \nu > 3/2,~\tau > 1 \} $&
$ a_1c_1 =\varsigma  $ \\
\hline
\hline
\end{tabular}
\end{table}

Obviously, the first condition (\ref{16}) is fulfilled if $a_1 =a_2 =a_3=0$. Calculating the first moment $m'_1(\varepsilon)$, one finds that the second limit (\ref{16}) leads to the condition   
\begin{equation}
\lim_{\varepsilon \to 0} \left\{ \varepsilon^{1-\mu}c_0
 \left[  {1 \over 2} \left( c_1 \varepsilon +{c_2 \over
\varsigma^2 } \varepsilon^{1 -\mu +\nu } \right)
 + {c_3 \over \varsigma }
 \varepsilon^{\tau}\right] \right\} = 1. 
\label{22}
\end{equation}
Using this limit, the third condition (\ref{16}) can easily be established by induction. Limiting equation (\ref{22}) determines another trihedral surface $S_{\delta'}$ 
(see figure~\ref{fig1}), which corresponds to the $\delta'$-limit. This 
surface is formed by the apex $Q_0$, the edges $Q_1, Q_2, Q_3$
and the planes $Q_{4}, Q_{5}, Q_{6}$ defined in table~\ref{tab:table2}.
Condition (\ref{22}) also imposes the constraints on the constants $c_0, c_1,
c_2, c_3$ and $\varsigma$ summarized in table~\ref{tab:table2}. 
\begin{table}
\caption{\label{tab:table2}
Definition of apex $Q_0$, edges $Q_1, \, Q_2, \, Q_3$ and planes 
$Q_{4}, \, Q_{5}, \, Q_{6}$ \\ forming the trihedral surface 
$S_{\delta'}$ together with the constraints 
on constants \\ $ c_0 \, c_1, \, c_2, \, c_3; \, \varsigma$, satisfying limit 
(\ref{22}). }
\begin{tabular}{ll}
\hline
\hline
$Q_j$-sets     & 
Constraints on $ c_0, \, c_1, \, c_2, \, c_3; \, \varsigma$\\
\hline 
 $Q_0 \doteq  \{ \mu = \nu = 2, ~ 
 \tau =1 \}$  &   ${1 \over 2} \left( c_1 + c_2  \varsigma^{-2} \right)
 + {c_3  \over \varsigma }   = {1 \over c_0} $\\
$Q_1  \doteq  \{ 1 < \mu < 2, ~ \nu =2(\mu -1),~
\tau =\mu -1 \} $ &   ${ c_2 \over 2\varsigma} + c_3 
= {\varsigma \over c_0} $ \\
 $Q_2 \doteq  \{  \mu =2, ~ \nu > 2,~
\tau = 1\} $ & $ { c_1 \over 2} + {c_3 \over \varsigma } 
 = {1 \over c_0} $ \\
$Q_3  \doteq  \{ \mu= \nu = 2,~
\tau > 1 \}$ & $  c_1+ c_2 \varsigma^{-2} = 
{2\over c_0} $   \\
 $Q_{4}  \doteq  \{ 1 <  \mu < 2, ~ 
\nu > 2(\mu -1) ,~ \tau =\mu -1 \} $ &  $c_0 c_3  = \varsigma $\\
$Q_{5}  \doteq  \{ 1 < \mu < 2,
~ \nu = 2(\mu -1),~
\tau > \mu -1 \} $ & $ c_0 c_2  = 2\varsigma^2  $ \\
 $Q_{6} \doteq  \{  \mu = 2, ~ \nu > 2,~\tau > 1 \} $& 
$c_0 c_1 = 2 $ \\
\hline
\hline
\end{tabular}
\end{table}

\section{A finite-range solution for the rectangular model }

The solution of equation (\ref{1}) can be written through the transfer matrix
$\Lambda$ connecting the boundary conditions for the wavefunction 
$\psi(x)$ and its derivative $\psi'(x)$ at $x=x_1=-l$ and $x=x_2 = \rho +r$: 
\begin{eqnarray}
\left( \begin{array}{cc} \psi(x_2)  \\
\psi'(x_2) \end{array} \right) 
 = \Lambda \left(
\begin{array}{cc} \psi(x_1)   \\
\psi'(x_1)   \end{array} \right), ~~~ \Lambda =  \left(
\begin{array}{cc} \Lambda_{11}~~ 
\Lambda_{12} \\
\Lambda_{21} ~~ \Lambda_{22} \end{array} \right) .
\label{23}
\end{eqnarray}
As a result, we obtain
\begin{eqnarray}
\Lambda_{11} &=& \left[
\cos(pl) \cos(qr) - {p \over q} \sin(pl) \sin(qr)\right]
\cos(s \rho) \nonumber \\
&-& \left[ {p \over s} \sin(pl) \cos(qr) + {s \over q}
\cos(pl) \sin(qr) \right]\sin(s \rho), \nonumber \\
\Lambda_{12} &=& \left[
{1 \over p} \sin(pl) \cos(qr) + {1 \over q} 
\cos(pl) \sin(qr) \right] \cos(s \rho) \nonumber \\
&+& \left[ {1 \over s } \cos(pl) \cos(qr) - {s\over pq}
\sin(pl) \sin(qr) \right]\sin(s \rho), \nonumber \\
\Lambda_{21} &=&  -\left[\, p \sin(pl) \cos(qr) + q \cos(pl) 
\sin(qr) \right]\cos(s \rho) \nonumber \\
&-& \left[ s \cos(pl) \cos(qr) - {pq \over s} \sin(pl) 
\sin(qr) \right]\sin(s \rho) , \nonumber \\
\Lambda_{22} &=& \left[
\cos(pl) \cos(qr) - {q \over p} \sin(pl) \sin(qr)\right]
\cos(s \rho) \nonumber \\
&-& \left[ {s \over p} \sin(pl) \cos(qr) + {q \over s}
\cos(pl) \sin(qr) \right]\sin(s \rho) ,
\label{24}
\end{eqnarray} 
where the quantities
\begin{equation}
 p \doteq \sqrt{E-\lambda h_1 }\, ,~~ q \doteq \sqrt{E-\lambda h_2 } \, ,~~  s \doteq \sqrt{E- \eta  h-\lambda h_3} 
\label{25}
\end{equation}
can be either real or imaginary.

\section{Asymptotical analysis: basic expansions}

The parametrization given by equations (\ref{18}) is a key point in our approach. Inserting these equations into (\ref{25}), in the $\varepsilon \to 0$ limit one can write the following asymptotics:
\begin{eqnarray}
&& p \to ~ \varepsilon^{1-\mu} \sqrt{-\lambda \left( a_1 
+ { c_0 \over c_1} \, \varepsilon^{\mu -2} \right) } \, ,
 ~~ ~ q  \to ~ \varepsilon^{1-\mu}  \sqrt{-\lambda 
\left( a_2 - {c_0 \over c_2} \, \varepsilon^{2\mu -2 -\nu}
\right) } \, , \nonumber \\
&& s \to ~ \varepsilon^{1-\mu} \sqrt{-\lambda a_3 -
(\eta/  c_3) \varepsilon^{2\mu -2-\tau}  }\, .
 \label{26}
\end{eqnarray}
Expanding the $\sin$- and $\cos$-expressions in (\ref{24}) up to the second order and using the asymptotics (\ref{26}) together with equations~(\ref{18}), one can calculate the asymptotics  of the matrix elements 
$\Lambda_{ij}=\Lambda_{ij}(\varepsilon),~i,j =1,2$, as $\varepsilon \to 0$. 
The element $\Lambda_{21}$ appears to be the most singular term in the region $\mu >1$ as $\varepsilon \to 0$. Both on the $S_\delta$ and $S_{\delta'}$ surfaces, it can be well defined only if an appropriate cancellation of singularities occurs in the $\varepsilon \to 0$ limit. Therefore, we start with the analysis of the expansion for this element, arranging the terms (given in powers of $\varepsilon$) in the series $\Lambda_{21} =
 \Lambda_{21}^{(0)} + \Lambda_{21}^{(1)}  +\ldots $, where 
 the group of terms $\Lambda_{21}^{(0)}$ contains divergences which under appropriate constraints cancel out in the $\varepsilon \to 0$ limit. Under these constraints appearing on $S_{\delta'}$ in the form of transcendental equations (called hereafter {\it transparency equations}), a non-zero transmission across the limiting zero-range potential 
$V(x)$ occurs. The next group $\Lambda_{21}^{(1)}$ contains the terms which appear to be finite either on the whole 
$S_\delta$ surface or in some non-empty subsets of 
 $S_{\delta'}$ if the transparency equations are taken into account. We denote this limit, which may be either zero or non-zero, as 
$\lim_{\varepsilon \to 0} \Lambda_{21}^{(1)} \doteq g$ and
the corresponding (transparency) subsets as 
$T_j \subset Q_j, ~j =0, \, 1, \ldots , \, 6$. 
The next terms of the expansion tend to zero on the sets
 as  $\varepsilon \to 0$. Using next the transparency equations, the other matrix elements either on $S_\delta$ or in $T_j$'s can be calculated explicitly. As a result, one finds that
$\lim_{\varepsilon \to 0} \Lambda_{12} =0$,
$\lim_{\varepsilon \to 0} \Lambda_{11} = \chi$ and 
$\lim_{\varepsilon \to 0} \Lambda_{22} = \chi^{-1}$
with a finite value $\chi$, so that the connection matrix in all the cases with a non-zero transmission takes the form
\begin{equation}
\Lambda = \left(\begin{array}{cc} \chi~~~0~~ \\
g~ ~~\chi^{-1}\! \!\end{array} \right).
\label{27}
\end{equation}
Particularly, as shown below, due to the cancellation of divergences (when $\mu > 1$) in the trihedral surrounded by the surface $S_\delta$ one obtains $\chi =1$ and $g = 0$, i.e. the full transmission, while on its boundary 
$S_\delta$, we have $\chi =1$ but $g \neq 0$ depending on the element $P_j, ~ j=0, \, 1, \, \ldots \, , 6$. Therefore, in the latter case $g$ may be called the coupling constant of an effective  $\delta$-interaction. Outside this trihedral, $ g \to \infty$, i.e. the potential 
$V(x)$ is fully nontransparent, except for the $T_j$-sets  where  $\chi \neq 1$ and  $g$ may be non-zero. 

 Because of the form of equations 
(\ref{24}), it is convenient to write the expansion for $\Lambda_{21}$ in the 
$\varepsilon \to 0$ limit separately for the following four cases:  (i) $pl \to 0$ and $qr \to 0$,
(ii) $pl \to 0$ but $qr$ tends to a non-zero finite constant,
(iii) $pl$ goes to a non-zero finite 
constant while $qr \to 0$, (iv) both $pl$ and $qr$ tend to non-zero finite constants. As follows from equations (\ref{18}) and asymptotics (\ref{26}), each of the four cases leads to certain constraints on $\mu$ and $\nu$ (see table~\ref{tab:table3}). 
 
\begin{table}
\caption{\label{tab:table3}
 Possible zero-range limits of $pl$ and $qr$ in $\sin$- \\
and $\cos$-expressions of transfer matrix solution (\ref{24}) \\
resulting in constraints on $\mu$ and $\nu$, 
and elements of surface $S_{\delta'}$ \\ where these limits are realized. } \, 
\begin{tabular}{lllll}
\hline
\hline
 &   $\mu$ & $\nu$ & $\tau =\mu-1$ &  $\tau \ge 2(\mu -1)$ \\
\hline
 (i)&  $1 < \mu < 2$ & $\nu >2(\mu -1)$ &  
$ Q_{4}$ &  $S_\delta $ \\
(ii) & $1 < \mu < 2$ & $\nu =2(\mu -1)$ & $Q_1$ & $Q_{5}$   \\
(iii) & $ \mu = 2 $ & $\nu >2$ & $Q_2$ & $ Q_{6}$\\
(iv)&  $ \mu =2$ & $\nu =2$ & $  Q_0 $ & $ Q_{3}$\\
\hline
\hline
\end{tabular}
\end{table}

We write the expansions in powers of $\varepsilon$ for each of cases (i), $\ldots$ , (iv),  separately, and then perform  
the explicit cancellation of divergences on each element of the surfaces
$S_\delta$ and $S_{\delta'}$. As summarized in table~\ref{tab:table1}, some elements of $S_{\delta'}$ lie in the $\tau=\mu-1$ plane, whereas the others including the surface $S_\delta$ belong to the $\tau \ge 2(\mu -1)$ half-space. In particular, for the $S_\delta$ surface we obtain the following expansion:
\begin{eqnarray}
&& \Lambda_{21}= \eta + \lambda c_0  \left( 1 - {1 \over \varsigma}  \right) 
\varepsilon^{1-\mu}   
+  c_1 \left[ \lambda a_1 + { \lambda^2 c_0^2 \over 2 } \left( {1 \over 3} - {1 \over \varsigma} \right)\right] \varepsilon^{3-2\mu}   \nonumber \\
&& + \, c_2 \left[ {\lambda a_2 \over \varsigma} 
+ { \lambda^2 c_0^2 \over 2 \varsigma^2 }
\left({1 \over 3\varsigma } -1 \right) \right]
 \varepsilon^{3-3\mu +\nu}  + c_3 \left(\lambda a_3 - {\lambda^2c_0^2 \over \varsigma} \right) \varepsilon^{2-2\mu +\tau} + \ldots . \nonumber \\
\label{28}
\end{eqnarray}
Here, the cancellation of divergences occurs only if 
$\varsigma =1$. As a result, 
 the coupling constant of the total $\delta$-interaction becomes
\begin{eqnarray}
g &=& \eta + \lambda \lim_{\varepsilon \to 0} 
\left[c_1 \left(a_1 - {\lambda \over 3} c_0^2 \right) 
\varepsilon^{3-2\mu } +c_2  \left(a_2 - 
{\lambda \over 3} c_0^2 \right) \varepsilon^{3-3\mu +\nu}
\right. \nonumber \\
&+& \left. c_3 \left(a_3 - \lambda  c_0^2 \right)  
\varepsilon^{2-2\mu + \tau}  \right] .
\label{29}
\end{eqnarray}
 Finally, using here the limit (\ref{21}), we obtain 
\begin{equation}
g = \eta +\lambda - { \lambda^2 c_0^2 \over 3}
\left\{ \begin{array}{lllllll}
c_1 + c_2 +3c_3 & \mbox{for} ~~ P_0 , \\
c_2 +3c_3 &  \mbox{for}~~ P_1 , \\
c_1 +3c_3 & \mbox{for}~~ P_2 , \\
c_1 +c_2 & \mbox{for}~~ P_3 , \\
3c_3 & \mbox{for}~~ P_{4} , \\
c_2 & \mbox{for}~~ P_{5} , \\
c_1  & \mbox{for}~~ P_{6}  .
\end{array} \right.
\label{30}
\end{equation}
All these constants are positive and arbitrary. The last term that depends on $P_j$ is a renormalization of the coupling constant $\lambda$. In the interior of the $S_\delta$ surface we have $g=0$ and outside this surface $g \to \infty$. Next, as follows from 
equations ~(\ref{24}), $\Lambda_{11}, ~\Lambda_{22} \to 1$ as $\varepsilon \to 0$, so that  
 $\chi=1$ on the $S_\delta$ surface including its interior.

Thus, the cancellation of divergences occurs both in the interior of $S_\delta$ (full transmission, $g=0$) and on its boundary resulting in an effective interaction of the 
 $\delta $-type. Therefore, the $S_\delta$ surface serves as a transition region from full to zero transmission.

\section{Transparency regimes on the $S_{\delta'}$ surface}

Similarly to the expansion (\ref{28}), the asymptotical analysis on the
$S_{\delta'}$ surface has to be performed separately on each element $Q_j, ~
j=0, \, 1, \ldots, 6$, of this surface. These seven cases with the corresponding
constraints on $\mu$ and $\nu$ are given in table~\ref{tab:table3}. Thus, 
 using equations ~(\ref{18}) and asymptotics (\ref{26}) with $a_1 =a_2 =a_3 =0$, and expanding the $\sin$- and $\cos$-expressions in equations (\ref{24}) up to the third order, we obtain the following series for cases (i), $\ldots$ , (iv):
$$
\Lambda_{21}(Q_4) =  \lambda c_0  \left( 1 - {1 \over \varsigma} - 
{\lambda c_0 c_3 \over \varsigma} \, \varepsilon^{1-\mu +\tau} \right) \varepsilon^{1-\mu} -  { \lambda^2 c_0^2 \over 2 }\left[ \left( {1 \over \varsigma} - 
{1 \over 3} + {\lambda c_0 c_3 \over 3\varsigma} \, \varepsilon^{1-\mu +\tau} 
\right)  \right. ~~~~~~~~~~~~~~~~~~~~~~
$$
\begin{eqnarray}
&\times &  c_1 \, \varepsilon^{3-2\mu }
+ \left. \left(1 -{1 \over 3\varsigma } - {\lambda c_0 c_3 \over 3 \varsigma} \, \varepsilon^{1-\mu +\tau} \right) {c_2 \over \varsigma^2 } \, 
\varepsilon^{3-3\mu +\nu}\right]  \nonumber \\
&+ & \, \eta\left[ 1 + { \lambda c_0 c_3 \over 2} 
\left( 1 - {1 \over \varsigma} \right)
 \varepsilon^{1-\mu +\tau}  -  {\lambda^2 c_0^2 c_3^2 \over 6 \varsigma} 
\varepsilon^{2(1-\mu +\tau)}\right] +   \ldots , 
\label{31}
\end{eqnarray}
$$ 
 \left( \cos \! { \sqrt{\lambda c_0 c_2}  \over \varsigma}
\right)^{\! \! -1} \Lambda_{21}(Q_1 \, \cup \, \, Q_5) = 
 \sqrt{ \lambda c_0 \over c_2} 
\left[ \sqrt{\lambda c_0 c_2} - \left( 1 + 
\lambda c_0 c_3 \, \varepsilon^{1-\mu + \tau } \right) \right.~~~~~~~~~~~~~~~~~~~~~~~~~~~~~~~~~~~~~~~~~~~~~
$$
\begin{eqnarray}
&\times & \left. \tan \!{ \sqrt{\lambda c_0 c_2} \over \varsigma}  \right] 
 \, \varepsilon^{1-\mu} 
  +  {\lambda c_0 c_1 \over 2 } \sqrt{ \lambda c_0 \over c_2} 
\left[ { \sqrt{\lambda c_0 c_2} \over 3 } - \left( 1 + {\lambda c_0 c_3 \over 3}
\varepsilon^{1-\mu +\tau} \right) \right. \nonumber \\
&\times &  \left. \tan \! { \sqrt{\lambda c_0 c_2}
\over \varsigma } \, \right] \varepsilon^{3-2\mu} 
+ \, \eta \left[ 1 + {\lambda c_0 c_3 \over 2}\varepsilon^{1-\mu + \tau } 
\right. \nonumber \\
& -&  \left.
{c_3 \over 2}\sqrt{\lambda c_0 \over c_2}\left(1 +  {\lambda c_0 c_3 \over 3}
\varepsilon^{1-\mu + \tau } \right)
\tan \! { \sqrt{\lambda c_0 c_2}\over \varsigma } \varepsilon^{1-\mu + \tau } 
  \right]+ \ldots ,
\label{32}
\end{eqnarray}
$$ 
\left( \cosh \! \sqrt{\lambda c_0 c_1} \, \right)^{\! \!-1}
\Lambda_{21}(Q_2 \, \cup \, \, Q_6) =  \sqrt{ \lambda c_0 \over c_1} 
  \left[ \left( 1 - {\lambda c_0 c_3 \over \varsigma } \varepsilon^{\tau -1}
 \right) \right. ~~~~~~~~~~~~~~~~~~~~~~~~~~~~~~~~~~~~~~~~~~~~~~~~~~~~~~~~~~~~~
$$
\begin{eqnarray}
& \times & \left. \tanh \! \sqrt{\lambda c_0 c_1} 
- {\sqrt{\lambda c_0 c_1} \over \varsigma} \right]  \varepsilon^{-1}
 + {\lambda c_0 c_2 \over 2 \varsigma^2 } \left[ {\lambda c_0 
\over 3\varsigma } \right. \nonumber \\
&- & \left. \sqrt{\lambda c_0 \over c_1} \left( 1 - 
 {\lambda c_0 c_3 \over 3 \varsigma } \varepsilon^{\tau -1} \right) 
\tanh \! \sqrt{\lambda c_0 c_1} \, \right] \varepsilon^{\nu -3} +
\eta \left[ 1  - {\lambda c_0 c_3 \over 2\varsigma } \varepsilon^{\tau -1}\right.
 \nonumber \\
& +& \left. {c_3 \over 2}\sqrt{\lambda c_0 \over c_1} \left( 1 - 
 {\lambda c_0 c_3 \over 3 \varsigma } \varepsilon^{\tau -1} \right) 
\tanh \! \sqrt{\lambda c_0 c_1} \, \varepsilon^{\tau -1} \right] 
+ \ldots  , \label{33}
\end{eqnarray}
$$
 \left( \cosh \! \sqrt{\lambda c_0 c_1} 
\cos{ \sqrt{\lambda c_0 c_2} \over \varsigma } 
\, \right)^{ \!\!  -1}\Lambda_{21}(Q_0 \, \cup \, \, Q_3) = \sqrt{ \lambda c_0 \over c_1 c_2} \left( \sqrt{c_2} \tanh \! \sqrt{\lambda c_0 c_1} \right.
~~~~~~~~~~~~~~~~~~~~~~~~~~~~~~~~~~~~~~~~~~
$$
\begin{eqnarray}
&-&  \left. \sqrt{c_1} \tan{ \sqrt{\lambda c_0 c_2} \over \varsigma } 
-  \sqrt{\lambda c_0 } \, c_3   \tanh \! \sqrt{\lambda c_0 c_1} \,
 \tan{ \sqrt{\lambda c_0 c_2} \over \varsigma } \, \varepsilon^{\tau -1} \right)  \varepsilon^{-1} \nonumber \\
&+& \eta \left[ 1 + {c_3 \over 2} \sqrt{\lambda c_0 \over 
c_1 c_2} \left( \sqrt{c_2} \tanh \! \sqrt{\lambda c_0 c_1} - \sqrt{c_1}
\tan{ \sqrt{\lambda c_0 c_2} \over \varsigma }\, \right)\varepsilon^{\tau -1} 
\right. \nonumber \\
&-&  \left. {\lambda c_0 c_3^2 \over 6 \sqrt{c_1 c_2} }
\tanh \! \sqrt{\lambda c_0 c_1} \tan \sqrt{ {\lambda c_0 c_2} \over \varsigma }
\, \varepsilon^{2(\tau -1)} \right] + \ldots ,
\label{34}
\end{eqnarray}
respectively. Each of these four series contains the group of three terms at the
singularity $\varepsilon^{1-\mu}$. There are two ways of cancellation of
divergences in this group. One of these can be performed in the $\tau = \mu-1$
plane, where the $Q_0, ~ Q_1, ~Q_2, ~Q_{4}$ elements are found (see also table~\ref{tab:table3}). All these three terms compose the 
$\Lambda_{21}^{(0)}$ group and participate in the cancellation. The other way occurs on the 
$Q_3, ~ Q_{5}, ~ Q_{6}$ elements being subsets of the 
$\tau \ge 2(\mu -1)$ half-space (table~\ref{tab:table3}). Here, the 
$\Lambda_{21}^{(0)}$ group consists of two terms which are to be canceled out. 

\begin{figure}
\centerline{\epsfig{file=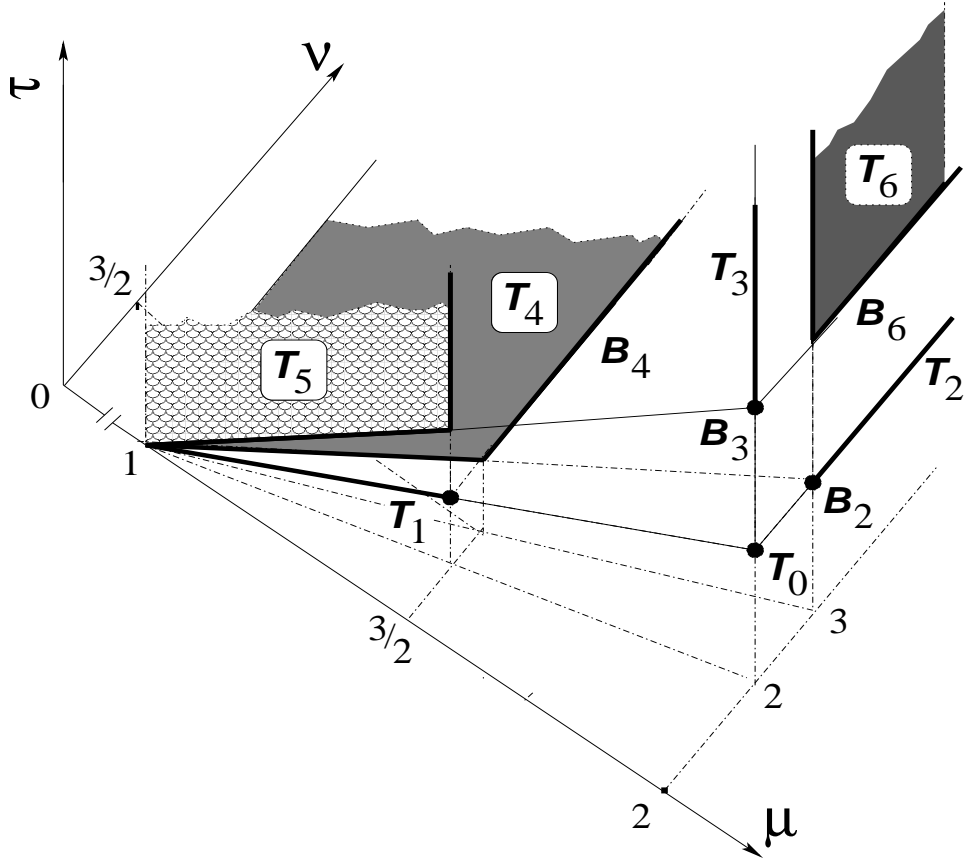,width=5.0 in,height=4.5 in,angle=0}}
\vspace{2pt}
\caption{ $T_j$-sets, $ j =0$ (point), 
$j= 1, \, 2, \, 3 $ (lines) and 
$j = 4, \, 5, \, 6$ (planes), together with their boundaries $B_j$, $j=1, \, 2, \, 3$ (points) and $j =4, \, 5, \, 6$ (lines) on the trihedral surface $S_{\delta'}$. Notations for $B_1$ and $B_5$ are omitted. Sets $B_1, \, B_2, \, B_3$ are shown by balls and sets $B_4, \, B_5, \, B_6$ by thick lines. }
\label{fig2}
\end{figure}

The cancellation of divergences imposes the constraints in the form of transparency equations (hereafter also called $T_j$-equations). Using these equations in the next terms of the expansion, one finds the subsets (called hereafter {\it transparency sets or $T_j$-sets}) 
$T_j \subset Q_j,~j=0, \, 1, \ldots, 6$, where the limit $g$ is finite. The
results of these calculations are summarized in table~\ref{tab:table4} for each element $Q_j$ of the $S_{\delta'}$ surface and illustrated by figure~\ref{fig2}. The calculation of the $\varepsilon \to 0$ limit of the other transfer matrix elements with taking into account the $T_j$-equations gives  representation (\ref{27}). 
\begin{table}
\caption{\label{tab:table4}
 Transparency equations and subsets of $S_{\delta'}$ given on each element $Q_j,$
\\ $j=0, \, 1, \ldots, 6$. }
\begin{tabular}{lll}
\hline
\hline
$Q_j$ & ~~~~$T_j$-equations & ~~~~~~~~~$T_j$-sets    \\
\hline 
\\
$Q_0$ ~~~~~~& $ { \sqrt{c_2} \, \tanh{\sqrt{\lambda c_0c_1}}
\over \sqrt{c_1} + \sqrt{\lambda c_0} \, 
c_3 \,  \tanh{\sqrt{\lambda c_0c_1}} } = 
\tan{ \sqrt{\lambda c_0c_2} \over \varsigma}$ 
&~~~ $T_0 \doteq \{ \mu = \nu = 2, \, \tau =1 \} = Q_0$ 
 \\ 
\\
$Q_1$ & ~~ $\tan \! { \sqrt{\lambda c_0 c_2} \over \varsigma } = 
{\sqrt{\lambda c_0 c_2} \over 1 + \lambda c_0 c_3 }$ &~~~
$T_{1} \doteq  \{ 1 < \mu \le 3/2, \, \nu = 2(\mu -1),$ \\
\medskip
 & & ~~~~$ \tau =\mu -1 \}$  \\
\medskip
$Q_2$ & ~~
 $  \tanh \! \sqrt{\lambda c_0 c_1} = {\sqrt{\lambda c_0 c_1} \over \varsigma - \lambda c_0 c_3  }$ &~~~
$ T_{2} \doteq \{ \mu = 2, \, \nu \ge 3, \, \tau =1 \}$
\\ 
\medskip
$Q_3$ &  ~~~~$  \sqrt{c_1} \, 
\tan{ \sqrt{\lambda c_0c_2} \over \varsigma} =\sqrt{c_2} \, \tanh{\sqrt{\lambda c_0c_1}} $  &~~~
$T_{3} \doteq \{ \mu = \nu = 2, \, \tau \ge 2 \}$  \\
\medskip
$ Q_{4}$ &~~~~ $ \varsigma (1-\lambda) =1 $ &~~~
$  T_{4} \doteq 
\{ 1 < \mu \le 3/2, \, \nu \ge 3(\mu -1), $\\
\medskip
& &~~~~ $ \tau =\mu -1 \} $ \\
\medskip
$Q_{5}$ & ~~~~$  \tan \! { \sqrt{ \lambda c_0 c_2 } \over \varsigma }= \sqrt{ \lambda c_0 c_2 } $ &~~~
$ T_{5}  \doteq 
\{ 1 < \mu \le 3/2, \, \nu = 2(\mu -1), $ \\
\medskip
&&~~~~ $\tau \ge 2(\mu -1) \} $ \\
\medskip
$Q_{6}$ &~~$
\varsigma \tanh \! \sqrt{ \lambda c_0 c_1 } = \sqrt{ \lambda c_0 c_1 } $ &~~~
$T_{6} \doteq \{ \mu =2, \, \nu \ge 3, \, 
\tau \ge 2 \} $ \\
\hline
\hline
\end{tabular}
\end{table}

\section{Reduced transparency equations and an inverse problem for non-zero transmission}

Thus, a non-zero transmission occurs under the transparency equations 
listed in table~\ref{tab:table4} and the constraints on the 
parameters $c_0, \, c_1, \, c_2, \, c_3$ and $\varsigma$ given 
in table~\ref{tab:table2}. Next, the direct way would be inserting these constraints into the transparency equations. However, the resulting equations are not
 sufficiently convenient for a further analysis. Therefore, we simplify them by 
 imposing some relations that couple $c_0, \, c_1, \, c_2, \, c_3$ and $\varsigma$ and do not contradict the constraints from 
table~\ref{tab:table2}. These relations are listed in table \ref{tab:table5} where 
for convenience of the corresponding transparency equation an 
additional parameter 
$b$ has been incorporated. In this table, the interval $(0, \infty)$ means 
that the corresponding parameter is positive and arbitrary; the dependence $b (\lambda ; \varsigma)$ or 
$c_0 (\lambda ; \varsigma)$ denotes that $b$ or $c_0$ is a solution of the corresponding 
transparency equation. The list of reduced $T_j$-equations together with
simplified expressions of $\chi$ and $g$ is given in 
 tables~\ref{tab:table6} and \ref{tab:table7}. As follows from table 
\ref{tab:table7}, for the case with $\eta =0$ the effective $\delta$-interaction 
appears only on the boundaries of the $T_j$-sets which are denoted by $B_j$, 
$j=1, 2, \ldots , 6$. These boundary sets are listed in table~\ref{tab:table8} and illustrated by figure~\ref{fig2}. The appearance of the $\delta$-interaction on
the $B_j$-sets is similar to that on the $S_\delta$ surface being the transition region from full to zero transmission.
 
\begin{table}
\caption{\label{tab:table5}
  Coupling between constants $b, \, c_0, \, c_1, \, c_2, \, c_3$ \\ and $\varsigma$  imposed for simplification of transparency equations. }
\begin{tabular}{lllllll}
\hline
\hline
$T_j$ & $\varsigma $ & $b $ & $ c_0$  & $c_1$ & $c_2$ & $c_3$ \\
\hline 
\smallskip
$T_0$ & $(0, \infty) $ & $ - $ & $c_0 (\lambda ; \varsigma)
$ & $ {1 \over 1+c_0} $ & $   { \varsigma^2 \over
1+c_0} $ & $ {\varsigma \over c_0 (1+c_0) }$ \\
\smallskip
 $T_1$ & $(0, \infty) $ & $ - $ & $c_0 (\lambda ; \varsigma)
$ & $ (0, \infty) $ & $  {2\varsigma^2 \over 1+c_0 }  $ &
 $ { \varsigma \over c_0 (1+c_0) } $ \\
\smallskip
 $T_2$ & $(0, \infty) $ & $ - $ & $c_0 (\lambda ; \varsigma)
$ &  $ {2 \over 1+c_0} $ & $  (0, \infty)  $ & $ {\varsigma \over c_0 (1+c_0)} $ \\
\smallskip
  $T_3$ &  $ (0, \infty) $ & $b(\lambda ; \varsigma)
$ & $ {2 \over (1+ b^{-1})c_1} $ & $ (0, \infty) $ & $ 
 {c_1 \over b}\varsigma^2    $ & $ (0, \infty) $ \\
\smallskip
$T_{4}$ & ${1 \over 1 -\lambda} $ & $-$ & $(0, \infty) $ &
$(0, \infty) $ & $(0, \infty) $ & $ 
{1 \over (1-\lambda) c_0 }$ \\
\smallskip
$T_{5}$ & ${ \tan \! \sqrt{2 \lambda} \over \sqrt{2 \lambda} }$& $-$ & 
$(0, \infty) $ & $(0, \infty) $ &$
{ \tan^2 \! \sqrt{2 \lambda} \over \lambda c_0}$ & $(0, \infty) $ \\
\smallskip
$T_{6}$ & $ { \sqrt{2 \lambda}\over 
 \tanh \! \sqrt{2 \lambda} } $ 
& $ - $ & $  (0, \infty)  $ & $ 2/c_0 $ & $  (0, \infty)$ & $  (0, \infty)$ \\ 
\hline
\hline
\end{tabular}
\end{table}
\begin{table}
\caption{\label{tab:table6}
Reduced transparency equations and \\ 
corresponding values of $\chi$ for connection matrix (\ref{27}).}
\begin{tabular}{lll}
\hline
\hline
$T_j$ & ~~~~~~Reduced $T_j$-equations & ~~~~~~$\chi $ \\
\hline
\\
\medskip
$T_0$ & ~~$
{  \tanh\sqrt{\lambda c_0 \over 1+c_0} \over {1 \over \varsigma} + 
\sqrt{\lambda \over  c_0(1 +c_0)} \, 
  \tanh\sqrt{\lambda c_0 \over 1+c_0}    } = \tan\sqrt{\lambda c_0 \over 1+c_0}  $ &~~~ $
 {\varsigma 
\sinh\sqrt{\lambda c_0 \over 1+ c_0} \over
\sin\sqrt{\lambda c_0 \over 1+ c_0} }  $ \\
\medskip
$T_1$ & ~~$
\left( {1 \over \varsigma}  + { \lambda \over 1+c_0}
 \right) \tan \! { \sqrt{2\lambda c_0 \over 1+c_0} } =
   \sqrt{2\lambda c_0 \over 1+c_0} $
& ~~~$ {1 + {\lambda \varsigma \over 1+c_0} \over 
\cos \! { \sqrt{2\lambda c_0 \over 1+c_0} } }$\\
\medskip
$T_2$ &~~ $  
\varsigma \left( 1 - {\lambda \over 1+c_0} \right)
 \tanh \! \sqrt{2\lambda c_0 \over 1 +c_0} = 
 \sqrt{2\lambda c_0 \over 1 +c_0} $&~~~
$ { \cosh \!  \sqrt{2\lambda c_0 \over 1+c_0 }
\over 1- \lambda/(1+c_0) } $\\
\medskip
$T_3$ &~~ $ {\varsigma \over \sqrt{b} } \, \tanh \! 
\sqrt{2\lambda \over 1+ b^{-1}} =
\tan \! \sqrt{2\lambda \over 1+b}  $&~~~$
 { \cosh \! \sqrt{2\lambda \over 1+b^{-1}} \over
\cos \! \sqrt{2\lambda \over 1+b} } $\\
\medskip
$T_{4}$ &~~~ $\varsigma (1- \lambda) = 1$ & ~~~
$~{1 \over 1- \lambda}$ \\
\medskip
$T_{5}$ & ~~~$  \tan \! \sqrt{2 \lambda} =\varsigma 
\sqrt{2 \lambda} $ & ~~~$ \left(\cos \!  \sqrt{2\lambda }
\right)^{-1} $\\
\medskip
$T_{6}$ &~~~ $\varsigma \tanh \! \sqrt{2 \lambda} = 
\sqrt{2 \lambda} $&~~~$ ~
\cosh \!  \sqrt{2\lambda } $\\ 
\hline
\hline
\end{tabular}
\end{table}
\begin{table}
\caption{\label{tab:table7}
Values of $g$ for connection matrix (\ref{27}) calculated \\
for each $T_j$-set. Here $\delta_{\alpha, \beta}$ stands for the Kronecker delta symbol. }
\begin{tabular}{ll}
\hline
\hline
$T_j$ & ~~~~~~~~~~~~~~~~~~~~~~~~~~~~~~~~~~~~~~~~ $g$  \\
\hline
\\ 
$T_0$ & ~~$
\eta \! \left[\cos\sqrt{\lambda c_0 \over 1+ c_0} \, 
\cosh\sqrt{\lambda c_0 \over 1+c_0}  + {\lambda \varsigma \over 3 c_0 (1 +c_0) } 
\sin\sqrt{\lambda c_0 \over 1+ c_0} \, \sinh\sqrt{\lambda c_0 \over 1+c_0}\, \right]
  $\\
\\
$T_1$ &~~ $
\left( 1 + {\lambda \varsigma \over 1 +c_0}\right)^{-1}
\left[ \eta \left( 1+ {\lambda^2 \varsigma^2 \over 3 (1 +c_0)^2 }\right)
- {\lambda^2 c_0^2 c_1 \over 3} \, \delta_{\mu , 3/2}\right]
\cos \! \sqrt{ 2 \lambda  c_0  \over 1 +c_0 }
$ \\
\\
$T_2$ & ~~$
\left( 1 - {\lambda \over 1+c_0 } \right)^{-1} \left[ \eta 
\left( 1 - {\lambda \over 1+c_0} + {\lambda^2 \over 3 (1+c_0)^2 }\right)
- {\lambda^2 c_0 \over 3\varsigma^2 (1 +c_0) } \, \delta_{\nu , 3} \right]
\cosh \! \sqrt{2\lambda c_0 \over 1+ c_0}
 $ \\
\\
$T_3$ &~~ $
\left[ 
\eta -  {2\lambda c_3 \over (1+b^{-1}) c_1^2 } \, 
\tanh^2 \! \sqrt{2\lambda \over 1 +b^{-1}}\,\,
\delta_{\tau, 2} \right] \cos\! \sqrt{2\lambda \over 1+b } \,
 \cosh\! \sqrt{2\lambda \over 1+b^{-1} } $ \\
\\
$T_{4}$ & ~~~$ 
\eta \! \left[ 1 + {\lambda^2 \over 3(1 -\lambda)} \right]
- {\lambda^2 c_0^2 \over 3} (1-\lambda ) \left[ c_1 \delta_{\mu , 3/2}
+ c_2 (1 -\lambda)\delta_{\nu , 3(\mu -1)} \right]
 $ \\
\\
$T_{5}$ & ~~~$ 
\left[ \eta - \lambda^2 c_0^2  \left( {c_1 \over 3}\, \delta_{\mu , 3/2} 
+ c_3 \, \delta_{\tau , 2(\mu -1)} \right) \right] \cos \! \sqrt{2\lambda }
$ \\
\\
$T_{6}$ & ~~~$ 
\left[ \eta - {\lambda c_0^2 \over 2 } \left( {c_2 \over 3
\sqrt{2 \lambda} } \, \tanh\! \sqrt{2 \lambda } \, \, 
 \delta_{\nu , 3}+c_3 \, \delta_{\tau , 2} \right) \tanh^2\! \sqrt{2 \lambda }
\right] \cosh \! \sqrt{2 \lambda }
 $ \\ \ \\
\hline
\hline
\end{tabular}
\end{table} 

As follows from table~\ref{tab:table6}, for continuous test functions
 ($\varsigma =1$) the $T_0$-equation is invariant under the transformation $\lambda \to -\lambda$. With this transformation the elements in the pairs 
($T_1, \, T_2$) and ($T_5, \, T_6$) are transposed. Next, the $T_3$-equation is invariant under the transformation $\lambda \to -\lambda, ~b \to b^{-1}$. In all these cases, we obtain the replacement $\chi \to \chi^{-1}$.  
The reduced apex and edge $T_j$-equations (with $j =0, \, 1,\,
 2, \, 3$) contain  the two parameters $\varsigma$ and $c_0$, whereas the plane $T_j$-equations (with $j =4, \, 5,\, 6$) only the parameter $\varsigma$. Therefore, in the former case one can tune two parameters to construct a regularizing sequence for a given $\lambda$. For instance, fixing $\varsigma$, i.e., a corresponding space of test functions
${\cal D}_{\varsigma}$,  one can find  $c_0$ as a function of $\lambda$. In the latter case, for 
each available $\lambda$ we have to fix a space ${\cal D}_{ \varsigma}$ according to the solution of the plane $T_j$-equations. These solutions 
$\varsigma = \varsigma (\lambda)$ are trivial (see the last three lines 
in table~\ref{tab:table6}). 

Finally, for any given $\varsigma >0$, the apex and edge $T_j$-equations 
can be solved numerically. The numerical solution of these equations 
for the case of continuous test functions ($\varsigma =1$) and $\lambda \in (0, \infty)$ is present in figure~\ref{fig3} where $j = 0, \, 1, \, 3$. Here, for any $\lambda$ exceeding some critical value $\lambda_{ c}$, there exists a countable sets of roots. 
Having the solutions for $\lambda$, one can calculate the matrix elements $\chi$
and $g$ according to the equations listed in tables~\ref{tab:table6} and \ref{tab:table7}, respectively. 

\section{Reflection-transmission coefficients and bound states}

Having the values for $\chi$ and $g$ (see tables~\ref{tab:table6} 
and \ref{tab:table7}), one can calculate the reflection-transmission coefficients and bound states 
through the elements of the matrix $\Lambda$ in the standard way~\cite{cnp,gnn,z_pla}. Indeed, using the definition for the reflection and transmission 
coefficients according to the equations
\begin{equation}
\psi (x) =  \left\{ \begin{array}{ll}
  {\rm e}^{{\rm i}k x} +R \, {\rm e}^{-{\rm i}kx}
 &  \mbox{for}~~~ - \infty < x < x_1 , \\
T \, {\rm e}^{{\rm i} k x}  & \mbox{for}~~~~~
  x_2 < x < \infty ,
\end{array} \right. ,
\label{35}
\end{equation}
 one can rewrite the following equations
\begin{eqnarray}
 \left( \begin{array}{cc} T \\
 {\rm i}k T \end{array} \right)  {\rm e}^{-{\rm i}k(x_1 -x_2)}
 = \left( \begin{array}{cc} \Lambda_{11}~~ \Lambda_{12} \\
\Lambda_{21} ~~ \Lambda_{22}\end{array} \right) \left(
\begin{array}{cc} 1+R  \\
 {\rm i}k (1-R)  \end{array} \right) .
\label{36}
\end{eqnarray}
Solving next this matrix equation with respect to the coefficients 
$R$ and $T$, one finds their representation in terms of the matrix elements $\Lambda_{ij}$: 
\begin{equation}
R  =  -  \, {\Lambda_{11} -  \Lambda_{22} + {\rm i} 
(k \Lambda_{12}  + k^{-1} \Lambda_{21}) \over \Delta}
 ~~~\mbox{and}~~~
T ={ 2 \over \Delta } {\rm e}^{- {\rm i} k x_0 } 
\label{37}
\end{equation}
where $\Delta \doteq  \Lambda_{11} +  \Lambda_{22}
 - {\rm i} ( k \Lambda_{12}  - k^{-1} \Lambda_{21} )$.
One can easily check the validity of the conservation law 
$|R|^2 + |T|^2 =1$.

Next we denote
\begin{equation}
u \doteq \Lambda_{11} - \Lambda_{22}~~\mbox{and}~~
 v \doteq k \Lambda_{12} + k^{-1}\Lambda_{21} .
\label{38}
\end{equation}
Then, we obtain with the use of the equality 
$ \Lambda_{11}  \Lambda_{22} - \Lambda_{12} \Lambda_{21}=1$
the following basic equations for the reflectibility and
transmissibility:
\begin{equation}
|R|^2 = { u^2 +v^2 \over 4+ u^2 +v^2 } ~~\mbox{and}~~
|T|^2 = { 4 \over 4+ u^2 +v^2 } .
\label{39}
\end{equation}
These coefficients can be rewritten in the form
\begin{equation}
|R|^2 = { ( \chi - \chi^{-1} )^2 + g^2/k^2 \over 4+ 
( \chi - \chi^{-1} )^2 +g^2/k^2 }\, , ~
|T|^2 = { 4 \over 4 + ( \chi - \chi^{-1} )^2 + g^2/k^2 } \, .
\label{40}
\end{equation}

In the case when $g \neq 0$, i.e. a $\delta$-potential is present, one can expect the existence of a nontrivial bound state with energy $E \doteq - \kappa^2$. Indeed, looking for negative-energy solutions of equation~(\ref{1}) in the form
\begin{equation}
\psi (x) =  \left\{ \begin{array}{ll} A \,
  {\rm e}^{\kappa x} 
 &  \mbox{for}~~~ - \infty < x < 0 , \\
B \, {\rm e}^{-\kappa x}  & \mbox{for}~~~~~
  0 < x < \infty ,
\end{array} \right. ,
\label{41}
\end{equation}
one can write the matrix equation
\begin{eqnarray}
 \left( \begin{array}{cc} B \\
 -\kappa B \end{array} \right)  
 = \left( \begin{array}{cc} \chi~~ 0 \\
g ~~ \chi^{-1}\end{array} \right) \left(
\begin{array}{cc} A \\
 \kappa A  \end{array} \right) .
\label{42}
\end{eqnarray}
The compatibility of solutions to this equation gives the equation for $\kappa$ from which we obtain 
\begin{equation}
\kappa = -  { g \over \chi + \chi^{-1}} .
\label{43}
\end{equation}
Thus, for a given $\lambda$ one can also calculate the bound level $\kappa$.
\begin{table}
\caption{\label{tab:table8}
Boundary sets $B_j \subset T_j, ~j =1, \, 2, \ldots , 6$: points $B_1, \, B_2, \, B_3$ \\ and lines  $B_4, \, B_5, \, B_6$.  }
\begin{tabular}{ll}
\hline
\hline
~~~~~~~~~~~~~~~~~~~~~~~~~~~$B_j$-sets   \\ 
\hline
$B_1 \doteq  \{  \mu = 3/2, \, \nu = 1, \, \tau =1/2 \}$  \\
$B_2 \doteq \{ \mu = 2, \, \nu = 3, \, \tau =1 \}$ \\
$B_3 \doteq \{ \mu = \nu =  \tau = 2 \}$ \\
$ B_{4} \doteq \{ 1 < \mu \le 3/2, \, \nu = 3(\mu -1), \, 
 \tau =\mu -1 \} \cup  \{ \mu = 3/2, \, \nu > 3/2, \, 
 \tau =1/2 \} $ \\
$B_{5} \doteq \{ 1 < \mu \le 3/2, \, \nu = 2(\mu -1), \, 
\tau = 2(\mu -1) \}\cup \{  \mu = 3/2, \, 
\nu = 1, \, \tau > 1 \} $  \\
$B_{6} \doteq \{ \mu =2, \, \nu \ge 3, \, 
\tau = 2 \} \cup \{ \mu =2, \, \nu = 3, \, \tau > 2 \}$ \\
\hline
\hline
\end{tabular}
\end{table}
\begin{figure}
\centerline{\epsfig{file=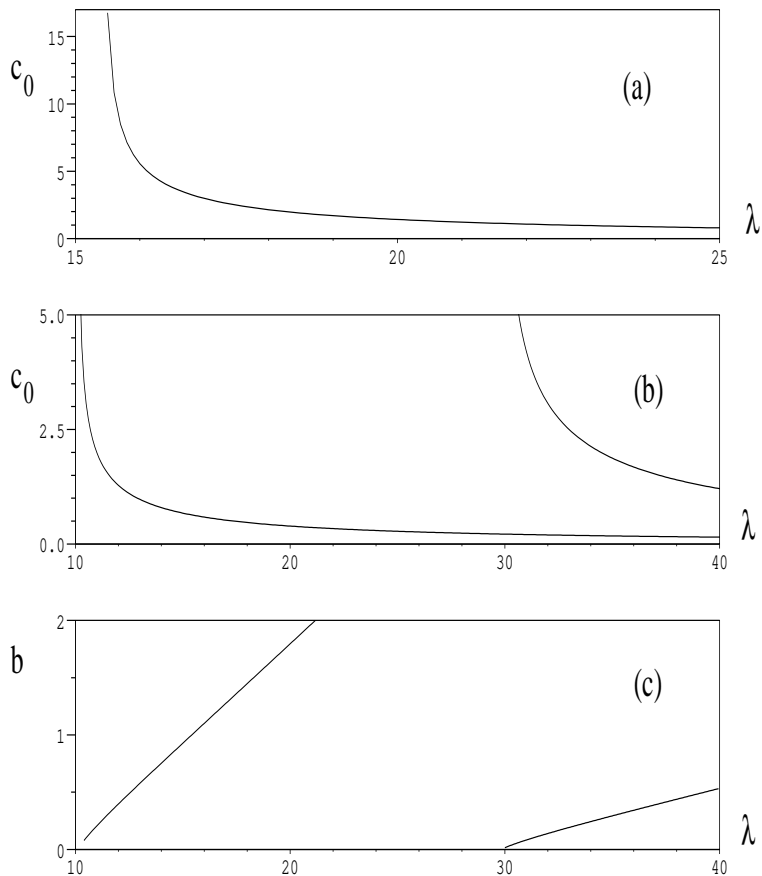,width=5.0 in,height=4.5 in,angle=-90}}
\vspace{2pt}
\caption{ Numerical solutions of the reduced $T_j$-equations (see 
table~\ref{tab:table6}) with
$j = 0, \, 1,  \, 3$ for continuous test functions ($\varsigma =1$): 
(a), (b) $c_0 = c_0(\lambda)$ and (c) $b =b (\lambda)$. The only 
first-root $\lambda$-dependence is plotted in (a) and the two 
first-root dependencies are plotted in (b) and (c). }
\label{fig3}
\end{figure}

\begin{figure}
\centerline{\epsfig{file=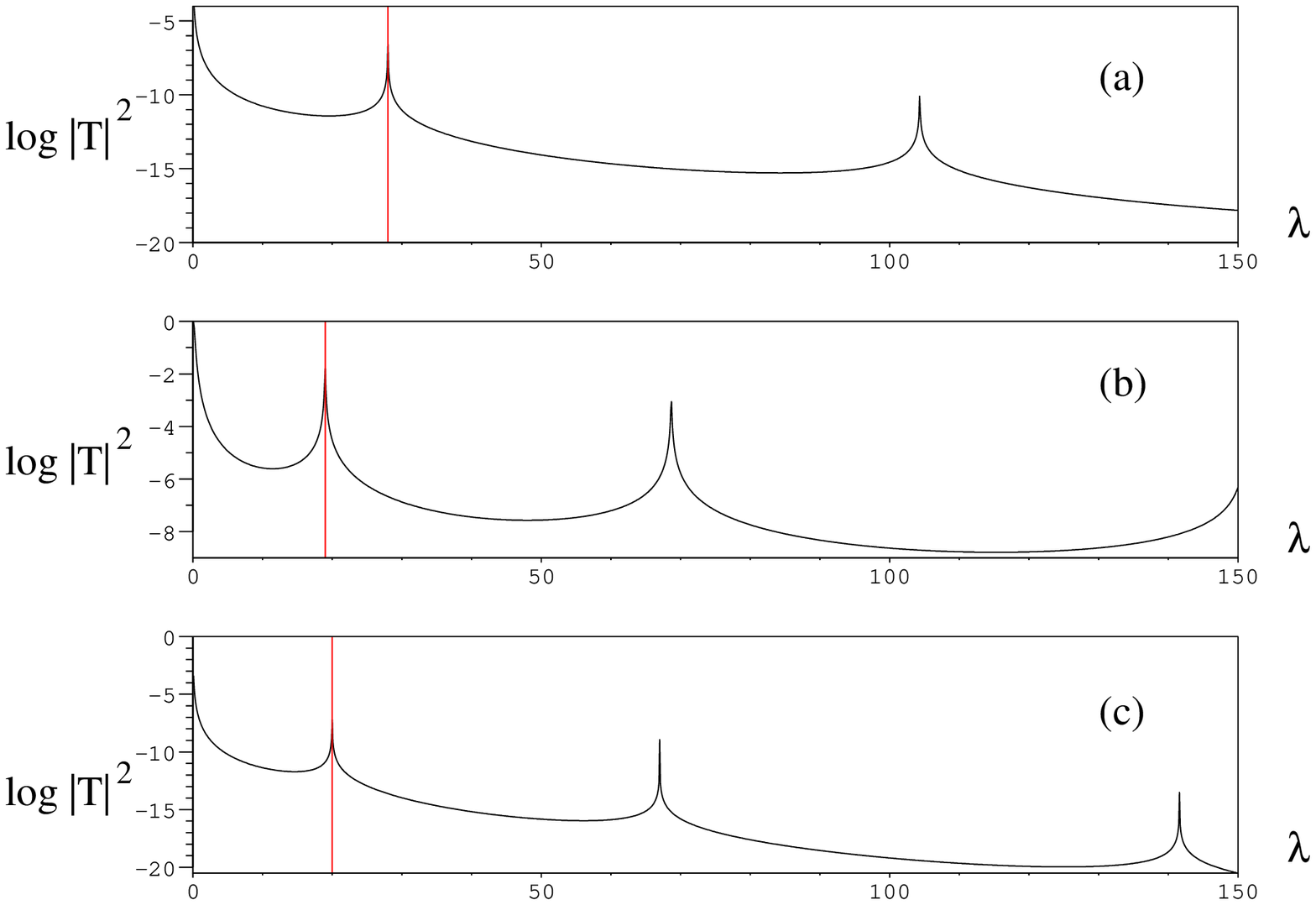,width=5.0 in,height=4.5 in,angle=0}}
\vspace{2pt}
\caption{(Color online) 
Resonant transmission for a given $\lambda =
\bar{\lambda}$ (shown by the vertical red lines) for three transparency sets
$T_0$, $T_1$ and $T_3$ calculated according to (\ref{24}), (\ref{38}) and
(\ref{39}) with parameters given in table~\ref{tab:table9}: (a) $T_0=Q_0$, $\bar{\lambda} = 28$, $c_0 = c_0(\bar{\lambda})$; (b) $T_1 = B_1$, $\bar{\lambda} = 19$, $c_0 = c_0(\bar{\lambda})$ and (c) $T_3 =B_3$, $\bar{\lambda} = 20$, 
$b  = b(\bar{\lambda})$. In all cases $\varsigma =1$,
 $\eta =0$, $k=1$, $c_1 =c_3 =1$ and $\varepsilon = 0.0001$.   } 
\label{fig4}
\end{figure}

As three examples, we have checked the inverse problem solution for the $T_0$-, $T_1$- and $T_3$-sets using the 
transfer matrix $\Lambda$ given by equations (\ref{24}) with sufficiently small $\varepsilon$. To this end, we solve numerically the 
corresponding $T_j$-equations from table~\ref{tab:table6}  with respect
 to $c_0$ and $b$, respectively, and thus obtain the multivalued 
functions $c_0=c_0(\lambda)$ and $b=b(\lambda)$. Next we fix any 
values $\lambda = \bar{\lambda}$ on each $T_0$-, $T_1$- and $T_3$-sets. 
It is sufficient to use only  the first roots of the $T_j$-equations. 
Inserting the values of $c_0(\bar{\lambda}$ and $b(\bar{\lambda}$ (see 
figure~\ref{fig3}) for given $\bar{\lambda}$'s into (\ref{18}) and (\ref{25}) and using table~\ref{tab:table5}, we obtain the corresponding expressions for the
parameters $l, \, r, \, \rho, \, p, \, q, \, s, $ which are summarized 
in table~\ref{tab:table9}. Here, the positive parameters $c_1$ and $c_3$ 
are arbitrary. Finally, inserting these parameters into equations 
(\ref{24}), (\ref{38}) and (\ref{39}), we plot the resonance behaviour of $|T|^2$ as a function of $\lambda$ as illustrated by figure~\ref{fig4}. As shown in this figure (see the red lines), the given $\bar{\lambda}$'s belong to the resonance sets 
$T_0, \, T_1$ and $T_3$. 
\begin{table}
\caption{\label{tab:table9}
Parameters $l, \, r, \, \rho, \, p, \, q, \, s$ used for calculation of trasmissibility $|T|^2$ according to 
equations~(\ref{24}), (\ref{38}) and (\ref{39}). Here parameters $c_0 = c_0(\bar{\lambda})$  and $b = b(\bar{\lambda})$ are fixed for a given $\bar{\lambda}$. Parameters $c_1$ and 
$c_3$ are arbitrary. }
\begin{tabular}{llll}
\hline
\hline
$T_j$ & ~~~ $l$~~~~~~& ~~$r$~~~~&~~~~$\rho$
~~~ \\
\hline
\\
$T_0$ &~~~$  {\varepsilon \over 1+c_0} $ &~ ${\varepsilon \over 1+c_0}$ & ~$  {\varepsilon \over c_0(1+c_0)}  $ \\
\\
$T_1$ &~ ~~$  c_1 \varepsilon $ & ~${ 2 \over 1 +c_0}
\varepsilon^{\mu -1} \, , ~~\mu \le 3/2 $ & ~$ 
{ 1 \over c_0(1 +c_0)}
\varepsilon^{\mu -1} \, , ~~\mu \le 3/2  $ \\
\\
$T_3$ &~ ~~$ c_1 \varepsilon $ &~ $  {c_1 \over b}
\varepsilon $ & ~$  c_3 \varepsilon^\tau \, , 
~~~~~~\tau \ge 2$ \\
 \\
\hline
\hline
$T_j$ & ~~~~~~~~ $p$~~~~~~& ~~~~~~~~~~~~~~~$q$~~~~~~~~&~~~~~~~~~~~~$s$
~~~~~~~ \\
\hline
\\
$T_0$ & ~$ {\rm i} \sqrt{\bar{\lambda} c_0 (1+c_0) \varepsilon^{-2} -k^2} $ &~ $\sqrt{ \bar{\lambda} c_0 (1+c_0) \varepsilon^{-2} +k^2}$ & ~$ {\rm i}
 \sqrt{\eta c_0 (1+c_0) \varepsilon^{-1} -k^2} $ \\
\\
$T_1$ &~ $ {\rm i} \sqrt{\bar{\lambda} (c_0 / c_1) 
\varepsilon^{-\mu} -k^2} $ & ~$\sqrt{\bar{\lambda} (c_0/2) (1+c_0) \varepsilon^{2(1-\mu)} +k^2}$ & ~$ {\rm i}
 \sqrt{\eta c_0 (1+c_0) \varepsilon^{1 -\mu} -k^2} $ \\
\\
$T_3$ &~ $ {\rm i} \sqrt{{ 2\bar{\lambda } b \over 1 + b} 
c_1^{-2}
\varepsilon^{-2} -k^2} $ &~ $\sqrt{ {2\bar{\lambda} b^2 
\over 1 + b}c_1^{-2}
\varepsilon^{-2} +k^2}$ & ~$  {\rm i} \sqrt{ {\eta \over c_3} \varepsilon^{-\tau} -k^2} $ \\ \ \\
\hline
\hline
\end{tabular}
\end{table}

\section{Conclusions and discussion}

In this paper, we have used a multi-parametric family of  regularizing sequences $ \Delta'_\varepsilon(x) $ of the $\delta'$-like shape (barrier-well rectangles with a squeezing parameter $\varepsilon$) to approximate both Dirac's delta function $\delta(x)$ and its derivative $\delta'(x)$.  A set of approximating parameters allows us to use a  `non-uniform squeezing' of rectangles instead of the standard `uniform squeezing' defined by the regularization (\ref{2}) with a dipole-like function ${\cal V}(\xi)$. In our approach, we have incorporated three powers $\mu, \, \nu, \, \tau$ and found in the  $\{ \mu, \, \nu, \, \tau \}$-space the two trihedral surfaces $S_\delta$ and $S_{\delta'}$ (illustrated by figure~1) that correspond to the distributional limits $ \Delta'_\varepsilon(x) \to \delta(x)$ and 
$ \Delta'_\varepsilon(x) \to \delta'(x)$ (as $\varepsilon \to 0$), respectively. Each triple point on $S_\delta$ or $S_{\delta'}$ determines a pathway along which the $\delta(x)$ or $\delta'(x)$ function is obtained. On the one hand, this regularization procedure generalizes pathway 
(\ref{2}), but on the other hand, it narrows the general class of dipole-like functions to the family of only piecewise functions. Since the limit $ \Delta'_\varepsilon(x) \to \delta(x)$ can be realized, it is worth to add in the Schr\"{o}dinger equation (\ref{1}) the $\delta$-potential with the corresponding approximation $ \Delta_\varepsilon(x) \to \delta(x)$.

Having explicit finite-range solutions of the Schr\"{o}dinger equation (\ref{1}) given through the transfer matrix 
(\ref{23})-(\ref{25}), one can control the cancellation of divergences that appear in the kinetic and potential energy terms along each pathway. This cancellation takes place for both the $\varepsilon \to 0$ limits:  $ \Delta'_\varepsilon(x) \to \delta(x)$ and $ \Delta'_\varepsilon(x) \to \delta'(x)$. In the $\delta$-case, the Schr\"{o}dinger operator is well defined on the whole surface $S_\delta$, whereas in the second case, it is defined only on some subsets of $S_{\delta'}$. This means that the existence or non-existence of the transparency regime depends on the pathway 
$ \Delta'_\varepsilon(x) \to \delta'(x)$. In particular, the different but correct scattering results: (i) the existence of resonance sets when regularizing by limit 
(\ref{2}) 
(proved in~\cite{c-g,gm,tn}) and (ii) and the zero transmission when using limit (\ref{4}) with $\alpha =1$ (calculated in \cite{pa}) can now be understood.   
It should be noticed here that the similar cancellation procedure has been realized to construct $\delta$-like point interactions in two~\cite{pc} and three~\cite{ca} dimensions as well as in the quantum field theory for the construction of non-trivial Hamiltonians for the Yukawa$_{1+1}$ and $\phi^4_{2 +1}$ interactions (for more details see~\cite{gj} an the comments with extensive bibliography on the constructive field theory given in~\cite{rs}). The similar problem arises also in the relativistic scattering by the $\delta$-potential~\cite{acg}. A three 
particle system with $\delta$-potentials in two and three dimensions 
has been studied in \cite{gs}. 

The main goal of this paper was to solve the following inverse problem: For a given $\bar{\lambda} \in \R\setminus{0}$ to construct a regularizing sequence $ \Delta'_\varepsilon(x) \to \delta'(x)$ as $\varepsilon \to 0$ such that 
$\bar{\lambda}$ would belong at least to one of the transparency sets $T_j$'s,
$j =0, \, 1, \ldots , 6$. To this end, we have developed an approach of
constructing the family of regularizing sequences that contains free parameters
as many as possible. In the $\delta'$-limit (when $a_1=a_2=a_3=0$), these are
the positive constants $\varsigma; \, c_0, \, c_1, \, c_2, \, c_3$ with some
constraints that provide the limiting distribution $\delta'(x)$. As illustrated
by table~\ref{tab:table4}, on each transparency set $T_j,~j=0, \, 1, \ldots ,
6,$ the $T_j$-equation for resonances has the same form depending only on these
constants. Taking into account the constraints given in table~\ref{tab:table2}, the number 
of these constants can be reduced and one of the ways of this reduction together
with an appropriate simplification is present in table~\ref{tab:table5}. The resulting simplified $T_j$-equations and the 
matrix elements $\chi$ and $g$ are given by tables~\ref{tab:table6} and~\ref{tab:table7}. On the sets $T_0, \, T_1, \, T_2, \, T_3$, except for $\varsigma >0$, it is possible to get out one parameter ($c_0$ or $b$) which can 
be tuned at fixed $\varsigma$ in each of the corresponding 
reduced $T_j$-equations. Solving these equations at a given $\lambda$, one 
can find numerically the dependence $c_0 = c_0 (\lambda; \varsigma)$ or $b = b (\lambda; \varsigma)$. If we fix $\varsigma =1$, the inverse problem can be solved for the space of continuous test functions. However, it is impossible to have such a parameter in the transparency equations on the sets $T_4, \, T_5, \, T_6$. Here, for a given $\lambda$ we have to fix $\varsigma$, i.e. the space of discontinuous test functions 
${\cal D}_\varsigma$. 

Finally, note that in spite of the Schr\"{o}dinger operators given 
by equation~(\ref{1}) and regularized by the 
sequence $\Delta'_\varepsilon(x) \to \delta(x)$ or
  $\Delta'_\varepsilon(x) \to \delta'(x)$ are well defined in 
  the $\varepsilon \to 0$ limit, the corresponding point interaction 
  models are ambiguous. From a physical point of view this means that 
  in some cases the exact description of a concrete point interaction 
  model has to be accompanied by a regularizing sequence.

 \bigskip

\noindent
{\bf Acknowledgments}
\bigskip

\noindent
We are grateful to Yu.D. Golovaty, V.P. Gusynin and 
I.V. Simenog for stimulating and helpful discussions. 

\bigskip

\noindent
{\bf References}

\bibliography{zz11}

\end{document}